\documentclass[10pt, conference, compsocconf]{IEEEtran}
\usepackage{cite}

 \usepackage[pdftex]{graphicx}
  \DeclareGraphicsExtensions{.pdf,.jpeg,.jpg,.png}
\usepackage{subfig}

\usepackage[cmex10]{amsmath}
\usepackage{algorithmic}
\usepackage{url}

\begin{document}
%
\title{Machine Science in Biomedicine: Practicalities, Pitfalls and Potential}


\author{\IEEEauthorblockN{T.W.  Kelsey}
\IEEEauthorblockA{School of Computer Science\\
University of St Andrews\\
St Andrews, United Kingdom\\
tom@cs.st-andrews.ac.uk}
\and
\IEEEauthorblockN{W.H.B. Wallace}
\IEEEauthorblockA{Department of Reproductive and Developmental Sciences\\
Division of Child Life and Health, University of Edinburgh\\
Edinburgh, United Kingdom\\
hamish.wallace@nhs.net}
}


%


\maketitle

\begin{abstract}
Machine Science, or Data-driven Research, is a new and interesting scientific methodology that uses advanced computational techniques to identify, retrieve,  classify and analyse data in order to generate  hypotheses and develop models.   In this paper we describe three recent biomedical Machine Science studies, and use these to assess the current state of the art with specific emphasis on data mining, data assessment, costs, limitations, skills and tool support.\end{abstract}

\begin{IEEEkeywords}
Text recognition; Data acquisition; Modeling; Biomedical computing;

\end{IEEEkeywords}

%
\IEEEpeerreviewmaketitle

\section{Introduction}
Machine Science \cite{Evans2010}, or Data-driven Research \cite{DDR}, is a new and interesting scientific methodology that uses advanced computational techniques to find, classify and analyse data in order to generate  hypotheses and develop models.   
It can be considered  a re-ordering of the classical approach to science, namely  (i) generate a hypothesis, (ii) design an experiment, (iii) collect data and (iv) analyse and discuss. In the modern world there are huge resources of organised scientific data, both in raw form and/or in summarised form.  For example, there is a long-standing requirement for all published microarray studies that the raw data is made available in a checkable repository \cite{DataRep}, proteomics research is moving towards a similar requirement \cite{Proteomics}, and most scientific papers report summaries of their data in charts and/or descriptive statistics.  Moreover, the explosion of knowledge in the form of digitised information has led to huge advances in computational tools and methods for searching, sorting and classifying data. Taken together, this motivates an alternative approach to science, namely (i) collect data, (ii) classify and organise the data, (iii) choose an analytic methodology based on the form, size and quality of the data, (iv) analyse and model, (v) report new patterns, uncovered inconsistencies and/or hypotheses that can be tested by future studies.
This approach is inherently multidisciplinary. The most important Computer Science skills and techniques involve artificial intelligence, data mining, machine learning, semantics \& ontologies, search, and modelling \& abstraction. Mathematical skills include non-linear regression, differential equations and statistical analysis. Without these, any Machine Science study is unlikely to succeed. However, an extremely good understanding  of the literature in the target science is needed: when combining datasets one has to be certain that they were collected in similar ways, even though the laboratory work may have been done on separate continents, in separate decades and using separate equipment. Therefore a high degree of professional understanding of the methods sections of studies reported in the literature is needed. 

Successful Machine Science studies have been published in the fields of neurology \cite{neurons2}, cancer research \cite{cancer1}, physics \cite{physics}, biophysics \cite{biophysics} and ecology \cite{ecology,basic}. Publications describing Machine Science tools and techniques for data extraction and management include \cite{databases1} and \cite{databases2}.

Machine Science is a hot topic in the theory and philosophy of modern science, with recent claims that 
\begin{quotation}
``within a decade, even more powerful tools
will enable automated, high-volume hypothesis
generation to guide high-throughput
experiments in biomedicine, chemistry, physics,
and even the social sciences''  \cite{Evans2010}. 
\end{quotation}

But, with specific focus on biomedicine,  what is the state of the art now? Is it possible to collect high quality raw data from existing studies? If not, can  data be re-created from published charts and descriptive statistics? To what extent can the obvious problems involving natural languages, scientific ontologies, entrenched scientific cultural viewpoints, homogeneity of data and standardisation of data from different studies be overcome?  Lastly -- but possibly most importantly -- how much does it cost in time, effort and equipment to produce high impact results in Machine Science. In this paper we partially answer these questions. We describe three recent biomedical Machine Science studies, and use these to (i) assess the current state of the art, (ii) identify the tools and techniques that currently support biomedical Machine Science, and (iii) highlight those areas which are likely to be resistant to successful full automation. The studies are all in the same scientific area: human reproductive  biomedicine. They differ in the availability and the quality of the data, all of which were obtained by careful aggregation of data from studies published in the biomedical literature. The specific results of the studies are unimportant: we focus on the practicalities and limitations of the scientific methodology employed, together with descriptions of useful tool support. 

The next three sections are detailed descriptions of the Machine Science studies, together with the hypotheses generated in each case.  In the final section we assess the strengths and weaknesses of current biomedical Machine Science, using both our exemplars and results from other Machine Science studies. 

\begin{figure*}[!t]
\centering
\includegraphics[width=6.2in]{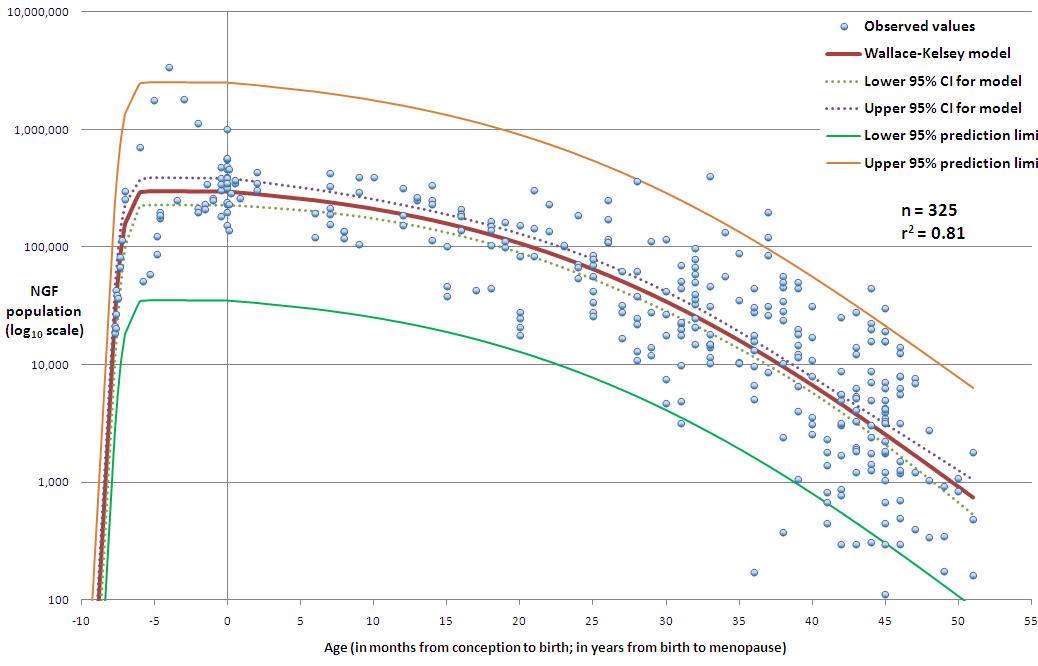}
 \caption{Model of ovarian reserve from mined biomedical data}
\label{WK}
\end{figure*}

\section{One homogeneous dataset}
\label{WKmodel}

We first describe a Machine Science study of human ovarian reserve from conception to menopause. Several studies have taken human ovaries (either after surgical removal or post-mortem); sliced, stained and mounted them; taken photographic images of the histological slides; and counted the number of non-growing follicles (NGFs) appearing in the images. In mammals, the NGF population defines ovarian reserve: once the population falls below about one thousand, ovulation ceases. The studies  typically involved subjects with restricted age ranges -- neo-natal, adolescent, fertile age -- and either reported the bare numbers (i.e. estimates of NGF populations) or produced models describing rates of decline of ovarian reserve that fitted their data and were applicable only within the age ranges of the studies. No model of ovarian reserve throughout life had been derived: the separate studies often compared their results to other studies, but reported only results based on   their data. 

The research question in biomedical Machine Science was: can these studies be identified and collated in order to produce a model of ovarian reserve that predicts (with explicit limits on accuracy) both the growth in NGF population after conception, and the subsequent decline in NGF population until menopause?  Derivation of such a model would be useful for comparison of the age-related dynamics of other indicators of human fertility, and could also lead to new hypotheses being generated for future investigation by reproductive biologists. 

Our research methodology was to systematically search the scientific literature for histological studies that estimated human ovarian NGF populations, then extract the data from these using automated or semi-automated techniques wherever possible. After assessing the studies for homogeneity, we automatically fitted 268 mathematical models to the combined dataset using TableCurve (Systat Software Inc., San Jose, California, USA), and selected as the gold-standard model the biologically plausible model with the highest coefficient of determination ($r^2$), Figure \ref{WK}.  

\subsection{Data mining}

We performed pubmed and medline searches, then followed the references in returned articles to identify older studies. Right from the start, the choice of search terms presents problems. NGFs can, and are, also referred to as primordial follicles in the literature. Moreover, the human oocyte is not the same biological entity as an NGF, but there is a one-one correspondence in their number so studies using this terminology have to be included.

 The strengths of the web-based search approach is that -- for recent publications -- the data mining process is largely automatic, with much of the non-intellectual work being performed by computer. PDF versions of the more recent studies -- together with full cross-referenced bibliographic information -- can be automatically downloaded into a cloud-based research tool for organising literature collections (we use Mendeley, but there are several others available). From the PDF it is possible to cut and paste tables into word-processing software (e.g. Microsoft Word), convert these into tables and export the result into a spreadsheet application (e.g. Microsoft Excel). 

Unfortunately, many of the studies were performed more than 25 years ago (Table \ref{Tdata}). This data had to be extracted by hand, since the journal articles had not yet been archived in digital form.  An extreme case was the 1951 PhD thesis by a Swedish researcher, which was obtained only by application in person at a medical library, followed by a wait of a few weeks for the thesis to be copied and mailed. This problem with older research remains, but many academic publishers are in the process of releasing digitised historical archives, so semi-automatic data mining should become easier in the future. 

It is not usual in biomedical literature to publish raw data in tabular form. The most common method of displaying data is in chart form with, for this study, age on the $x$-axis and NGF population on the $y$-axis. The extraction of individual datapoints is now semi-automated in the sense that software (such as PlotDigitiser) can be used to calibrate the axes, so that clicking on a datapoint entry loads an accurate estimate of that datapoint into a spreadsheet. Accuracy depends on the quality of the images of the charts, but in our experience the inter- and intra-observer coefficients of variation are very small (typically less than 2\%). Repeated datapoints are problematic: the study reports $n$ subjects, but there appears to be fewer points in the describing chart. Often the missing points can be deduced by analysis of the describing statistics for the data (e.g. median, minimum \& maximum; or mean $\pm$ one standard deviation).

\begin{table}[!ht]
\caption{
The eight quantitative histological studies forming the combined dataset}
\begin{tabular}{|lc|cccc|}
        \hline \multicolumn{2}{|c|}{Study} &  \multicolumn{4}{|c|}{Statistics}\\ 
         First author  & Year & Ovaries & Min. age & Max. age & Median age\\ \hline \hline
         Bendsen & 2006 & 11 & -0.6 & -0.6 & -0.6 \\ 
          Baker & 1963 & 11 & -0.6 & 7.0 & -0.2 \\ 
          Forabosco & 2007 & 15 & -0.5 & 0.5 & -0.3 \\ 
         Block & 1953 & 19 & -0.2 & 0.0 & 0.0 \\ 
          Hansen & 2008 & 122 & 0.1 & 51.0 & 38.0 \\ 
         Block & 1951 & 86 & 6.0 & 44.0 & 28.0 \\ 
         Gougeon & 1987 & 52 & 25.0 & 46.0 & 39.5 \\ 
          Richardson & 1987 & 9 & 45.0 & 51.0 & 46.0 \\ \hline \cline{3-6}
         \multicolumn{2}{|c|}{Overall} & 325 & -0.6 & 51.0 & 32.0 \\ \hline
      \end{tabular} 
\label{Tdata}
 \end{table}

\subsection{Data assessment}
This stage involves the most multidisciplinary expertise: can data from different studies be considered homogenous in terms of (a) biomedical laboratory practice, (b) standard levels of error, and (c) alignment with the Machine Science study in question? The answer to these questions is for the most part subjective, and will depend heavily on the skills and experience of individual investigators. However, it should be noted that -- for many biomedical studies -- the errors in the data are higher than any errors induced by the data mining process. There is no published standard error for histological  NGF population estimates, but it is likely to be at least 10\%. Data-mining errors are an order of magnitude lower. 

\subsection{Results and hypotheses}
Automated non-linear regression (or curve-fitting) is a very powerful tool. Rather than selecting models in advance and calculating goodness-of-fit, it is now possible to fit hundreds of models to see which best fits the data. Our model  \cite{Wallace2010} shows for the first time that for 95\% of women by the age of 30 years
only 12\% of their ovarian reserve is present and by the age of 40 years only 3\% remains. Further analysis of the model shows that the number of NGFs lost per month peaks at age 14--15. This provides the hypotheses that (a) changes in levels of NGF loss lead to puberty, (b) hormonal pubertal factors lead to changes in NGF loss, (c) the peak at this age is a fluke and there is no causal relation, and (d) our model is simply wrong, with peak loss actually occurring at some other age on average.  Another hypothesis generated by this study is that variations in NGF populations are more related to age for younger women, with lifestyle and genetic factors coming more into play later in life. 

The key point is that these hypotheses are now testable by biomedical researchers. There is a solid framework for designing experiments in terms of event size, cohort size needed, predicted outcomes, and confidence intervals. The Machine Science study may well lead to a greater understanding of reproductive biomedicine by motivating studies that would otherwise never be planned or undertaken.

\subsection{Costs, limitations and caveats}
 
This study was very cost efficient. The only non-human expenditure involved was computer connected to the internet,  software costing a few hundred US dollars in license fees, and small amounts of travel money for collaborative research and editorial meetings. To perform one of the studies that comprise our combined dataset, one would need a fully equipped histology lab (with scanners, microscopes, stains and ovens costing thousands of US dollars each), plus the labour costs associated with obtaining ovaries, fixing, slicing, staining and mounting them.  In terms of academic effort, again the Machine Science study was relatively inexpensive: much of the data acquisition was either automated or semi-automated, as was much of the mathematical and statistical analysis of the data. It takes many hours to count NGFs appearing in histology images, with more than one count needed to guard against individual mistakes. 

The main limitation of the study is that the model is not validated in any way. With only 325 datapoints covering over 50 years of age, we are relying on future studies to validate (or otherwise) the model or otherwise.  Another obvious limitation is that important data sources may have been missed in the systematic search. This is particularly true of research output published in languages other than English; these papers are unlikely ever to be found by the search procedure used, and would have to be interpreted by a biomedical expert conversant with the publishing language. 

Even though large parts of the research effort was performed by computers, it simply isn't possible to fully automate this type of the study. Some of the data sources are (in modern scientific terms) archaic, each paper has to be read and interpreted by at least one expert human in order to include or exclude its contents from the study.

\section{Two similar datasets: Machine Learning}

\begin{figure*}[!t]
\centerline{\subfloat[Case I]{\includegraphics[width=3.5in]{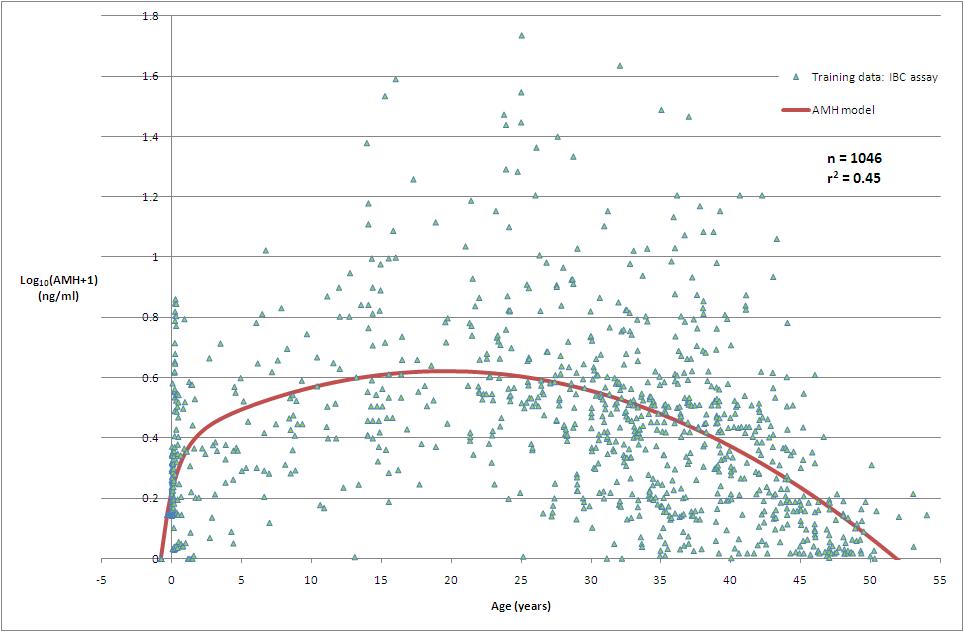}%
\label{fig_first_case}}
\hfil
\subfloat[Case II]{\includegraphics[width=3.5in]{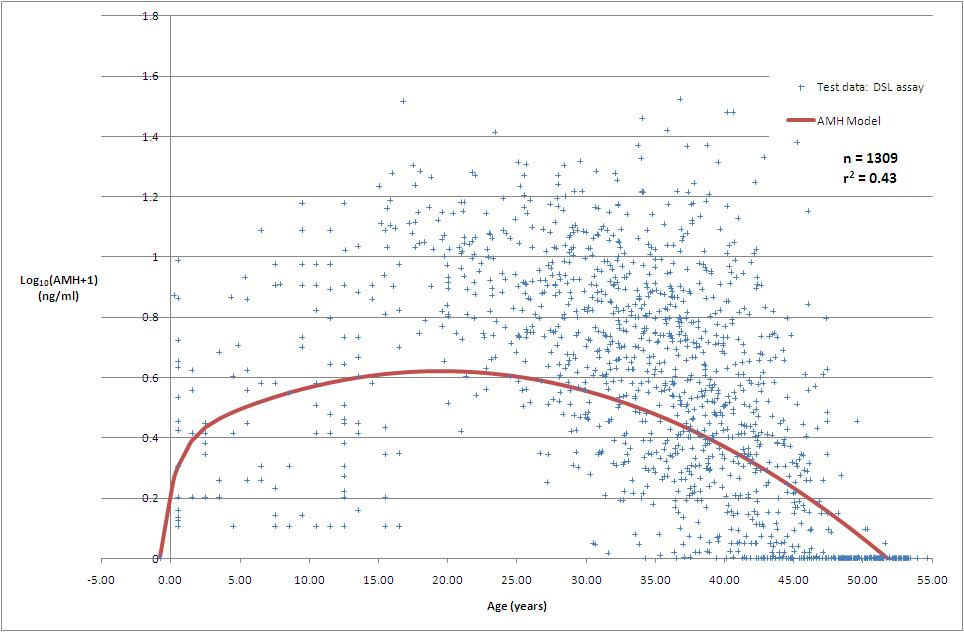}%
\label{fig_second_case}}}
\caption{Serum AMH model: traing data \& test data}
\label{fig_sim}
\end{figure*}

In this section we describe a similar study to the ovarian reserve model described in Section \ref{WKmodel}. The key difference is that we demonstrate that it is possible to overcome the lack of model validation by using machine learning techniques. The study involves the collection of data that related serum Anti-M\"{u}llerian Hormone (AMH) levels to chronological age. AMH is known to be related to human fertility, as it is produced by granulosa cells in the ovary. It is known that AMH levels are  below detection limits for post-menopausal women, and are elevated in women with polycystic ovarian system. As with ovarian reserve, no model has been published that gives average AMH levels for all ages, together with well-defined upper and lower prediction limits.

\subsection{Data mining}
As before, a web-based search of the biomedical literature was performed, taking great care to check for both for different but synonymous labels, and also similarly-defined but different data. AMH and MIS are synonyms, and nm/ml and $\mu$m/l represent the same measure. However, there are two different commercial assays used to measure AMH levels in blood, and these will return different values when used on the same sample. It is estimated that one gives values roughly twice those of the other, but the true conversion formula is probably hyperbolic and is simply not known. 

We therefore separated the data by assay used. AMH levels are easy, but expensive, to obtain in the laboratory, and we found about two thousand datapoints. Again, wherever possible, tabular data was imported directly into spreadsheets, but in the general case data had to be extracted from charts. In this case we also had access to raw data from an as yet unpublished study.

\subsection{Data assessment}
This was, again, the most subjective and multidisciplinary stage. Defining inclusion and exclusion criteria that ensured that our   study cohort could be assumed to approximate a normal fertile female population involved the identification of control subjects from comparative studies and healthy subjects from longitudinal studies. This process was painstaking and manual, as was the careful analysis of standardisation of studies in terms of assays,  units, coefficients of variation and detection limits.

\subsection{Results and hypotheses}
We collected two  datasets of similar size and having similar descriptive statistics (mean, median, max., min., and SD). We fitted curves to one set -- the training data -- and selected by highest coefficient of determination as before,  obtaining an $r^2$ of 0.45 (Figure \ref{fig_first_case}). We then calculated the residuals of the test data from this model, giving an  $r^2$ of 0.43 (Figure \ref{fig_second_case}). This 95\% agreement in goodness-of-fit for unseen data shows that the model can be used to accurately describe independent serum AMH levels, and is therefore highly likely to be correct in biomedical terms. 

Comparison of the ovarian reserve and AMH models suggests that ovarian activity (number of NGFs lost) correlates either well or extremely well with levels of AMH in the blood (birth to menopause correlation $r = 0.81$; birth to age 20 $r = 0.95$). However NGF population (or ovarian reserve) correlates much less well ($r = -0.88$ and $r = 0.38$ for the same age ranges). We can therefore hypothesise that AMH and NGF loss are biologically related in some way as yet unknown, especially early in life. As before, this hypothesis can now be tested using the results of the Machine Science study to design and assess the experiments.

\subsection{Costs, limitations and caveats}

This study cost a small fraction of the projected cost of a laboratory-based study to answer the same research question. We did not have to recruit 2,400 subjects, take blood samples or measure hormone levels. Our main cost was in time taken to ensure that the data collected was suitable as the basis for a classical machine learning modelling approach. Moreover, for this study most of the data came from recent (less than ten years old) literature, making the data mining process much simpler. 

By using machine learning to validate the AMH model, it is our opinion that the limitations and caveats associated with the ovarian reserve study no longer apply. This model has been independently validated and can hence be considered as the gold-standard for this hormone in female human blood.

\section{Data from descriptive statistics}

Ovarian volumes are another potential indirect indicator of human reproductive capability \cite{WK-04}. Volumes are easily obtainable using transvaginal ultrasound, with a standard error of 10\% or 12\% for 2- and 3-dimensional ultrasound respectively \cite{Brett}.

\subsection{Data mining}
Several studies have recorded volumes at known ages, but the data is dominated by a single study involving 58,673 observations of 13,963 subjects aged 25--91 years \cite{Pavlik2000}. Any Machine Science study that did not include this data would be seriously flawed, but the raw data is not available from the research group that owns it. For obvious reasons they did not produce a scatter plot containing 58,673 entries. The published summary of the data consists of a chart with entries for age, number of subjects of that age, mean volume and the upper 95\% prediction limit for that age. If volumes were normally distributed, it would be simple to extract the standard deviation from the prediction limit, then use a statistical software package (such as R) to randomly create a dataset containing the correct number of entries for each age, with the distribution for each age satisfying the pre-calculated normal variance. The resulting dataset would not be identical to that obtained by the original study, but would be statistically indistinguishable. Given the 12\% standard errors in the original data, we claim that the artificial dataset is an accurate representation of ovarian volumes for that many women of those ages. 
For lognormally distributed data the situation is slightly harder. Given the published mean and standard deviation, $\mu$ and $\sigma$, we can calculate their lognormal equivalents, $x$ and $y$, by solving the equations
$$
\mu = e^{x + \frac{1}{2}y^2} 
$$
$$
\sigma = e^{2x + y^2}(e^{y^2} - 1).
$$ 

We can then randomly create as many full datasets with the same statistical properties as is needed to produce a reliable model.

\subsection{Data assessment}
Is it sound scientific practice to recreate datasets from partial information published in the biomedical literature? We believe so, provided that the methodology used is explicitly stated. The key strength, in terms of modelling, is that many such  datasets can be constructed, so that the validity and robustness of any models produced can be exhaustively analysed. 

\subsection{Results and hypotheses}
Our results show both that ovarian reserve correlates well with ovarian volume \cite{WK-04}, although measurement errors reduce the efficacy of this indirect test of ovarian reserve. The hypothesis is that a combination of easily measured indirect factors will produce a reliable test for the number of years of reproductive life remaining for individual women. 

\subsection{Costs, limitations and tool support}
The cost of taking 58,673 observations was immense. The cost of using the results of this effort to produce comparative models was very small. The tool support needed is a textual data-mining system, a PDF to spreadsheet system, a computer algebra system to solve the lognormal equations, and a statistical analysis system to create datasets.

\section{Conclusion}

We have described three Machine Science studies, each based on mined data from the biomedical literature. Our primary conclusion is that biomedical Machine Science is both cost-effective and effective, producing the first ever models for important human biomedical attributes. High quality tool support is easily available, and more is being developed every year. The quantity and the quality of retrievable data is also increasing with time. There are strong bi-directional  links with the more usual  (often laboratory-based) biomedical scientific approach:
our Machine Science is almost entirely based on published data obtained from laboratory studies, and our hypotheses provide laboratory research groups with new ideas for future experimental directions. 

Despite these benefits, pitfalls remain. Without a high level of expertise in the subject area (specifically the ability to understand and assess the minutiae contained in methods sections of biomedical journal papers), successful Machine Science is unlikely. It should also be remembered that important data and information is contained only in paper form, and until and unless all the scientific literature is both digitised and searchable these important data can easily be missed. 

In summary, biomedical Machine Science is far from easy and requires expert multidisciplinary collaboration, but it leads to new insights, models and hypotheses that laboratory based research is highly unlikely to uncover.


\section*{Acknowledgment}

TWK is supported by United Kingdom Engineering and Physical Sciences Research Council (EPSRC) grants EP/CS23229/1 and EP/H004092/1. The
funders had no role in study design, data collection and analysis, decision to publish, or preparation of the manuscript.



\bibliographystyle{IEEEtran}
\bibliography{KW}
%


%


\end{document}